\documentclass[twocolumn,prb,showpacs,superscriptaddress]{revtex4}

\usepackage{graphicx}
\usepackage{dcolumn}
\usepackage{bm}

\input epsf.tex

\begin{document}

\title{Physisorption of positronium on quartz surfaces}

\author{R. Saniz}
\affiliation{Department of Physics and Astronomy,
Northwestern University, Evanston, Illinois 60208-3112, USA}
\author{B. Barbiellini}
\affiliation{Department of Physics,
Northeastern University, Boston, Massachusetts 02115, USA}
\author{P. M. Platzman}
\affiliation{Bell Laboratories, Lucent Technologies,
600 Mountain Avenue, Murray Hill, New Jersey 07974, USA}
\author{A. J. Freeman}
\affiliation{Department of Physics and Astronomy,
Northwestern University, Evanston, Illinois 60208-3112, USA}

\date{\today}
                                                                
                                             
\begin{abstract}

The possibility of having positronium (Ps) physisorbed at a material
surface is of great fundamental interest, since it can lead to new
insight regarding quantum sticking and is a necessary
first step to try to obtain a Ps$_2$ molecule on a material host.
Some experiments in the past have produced
evidence for physisorbed Ps on a quartz surface,
but firm theoretical support
for such a conclusion was lacking.
We present a first-principles density-functional calculation of the key
parameters determining the interaction potential
between Ps and an $\alpha$-quartz surface.
We show that
there is indeed a bound state with an energy of
$0.14$ eV, a value which agrees very well with
the experimental estimate of $\sim0.15$ eV.
Further, a brief energy analysis invoking the Langmuir-Hinshelwood
mechanism for the reaction of physisorbed atoms shows that the
formation and desorption of a Ps$_2$ molecule in that picture
is consistent with the above results.

\end{abstract}

\maketitle

Since positronium was first observed in 1951,\cite{deutsch51}
positronium physics has
evolved into a vigorous and fascinating field, at the intersection of
nuclear, atomic and condensed matter physics.\cite{surko01}
The positronium (Ps) atom,
i.e., an electron-positron bound state, is rather unique in that it
is one of the very few examples of exotic particle-antiparticle atoms
that have been observed, and the only one in the lepton family
(the other examples, such as charmonium, being in the hadron family).
Ps is produced by irradiating matter with positrons, and its formation and
stability has been studied in a variety of solid
hosts.\cite{schultz88}
In the case of metals, the injected positrons thermalize and can
reach the surface, where they can bind with an electron to form Ps.
Thus, Ps desorbed from Cu and Al surfaces has been well measured
and characterized.\cite{mills79,mills91}
In the case of insulators, in which Ps can also be formed in the bulk,
Ps emission from Al$_2$O$_3$, SiO$_2$, and MgO powders was reported
even earlier.\cite{paulin68,curry71}

A question of great interest in these systems is whether
Ps can be bound to the surface of the host. Indeed, this is related to
fundamental questions such as the nature of quantum
sticking\cite{mills91} and the potential of a Ps trap to lead to
dipositronium (Ps$_2$) formation and its connection to the realization
of a Ps Bose-Einstein condensate.\cite{surko01,platzman94}
In 1976, intriguingly low decay rates in
precision Ps lifetime measurements in SiO$_2$
powders were reported.\cite{gidley76} These findings were analyzed
theoretically considering the possibility of Ps physisorption on the
surface of the grains,\cite{ford76} but the conclusion was that no
surface bound state was possible.
Later, however, temperature-dependent Ps emission
measurements produced
experimental evidence suggesting Ps physisorption on an
$\alpha$-quartz single crystal surface,\cite{sferlazzo85} with
an estimated binding energy of $\sim0.15$ eV.
Very recently, moreover,
Ps lifetime measurements of high-density Ps
gas in nanoporous silica
suggest the formation of Ps$_2$ molecules, a conclusion assuming
that Ps atoms are previously 
physisorbed on the surface of the nanopores.\cite{cassidy05}

Consequently, we think that it is important to reconsider the
possibility of Ps physisorption on quartz surfaces.
In the case of Al surfaces, for instance, the longer than
expected (again) positron
lifetimes observed in experiment\cite{lynn84} can be understood
in terms
of a model calculation\cite{platzman86} of weakly physisorbed
Ps on the metal surface.
Although an alternative explanation invoking trapped positrons
in a slowly varying
inhomogeneous electron gas was proposed
by other authors,\cite{puska94} the former possibility
still exists.\cite{comment0}
Moreover, it is important
to recognize that the previous theoretical estimate
mentioned was only approximate.\cite{ford76}
It is clear that a more accurate calculation of the Ps-quartz
surface interaction potential is essential
in order to discard or support
the Ps physisorption interpretation of the above experiments
and to
find reference values for the physical parameters
characterizing the Ps-surface interaction.
Atomic physisorption on single-crystal surfaces has been extensively
studied in the past. An account of the typical values found in
experiment several characteristic parameters can be found in
Ref.~\onlinecite{hoinkes80}. From the theoretical point of view,
following the pioneering work of Zaremba and Kohn,\cite{zaremba76}
a good calculational scheme emerged thanks to the contribution
of several authors,\cite{nordlander84} which we essentially follow
here. In this respect,
of foremost importance are accurate calculations of the
dielectric function of the crystal and of the electronic charge
density profile at its surface.
In this work, we present a density-functional
calculation of these two quantities.
Our results for the Ps/$\alpha$-quartz surface interaction potential
show that indeed Ps has surface bound states. As will be
seen below, we find that
the ground state binding energy is very close to the experimentally
suggested value, which is also consistent with the possibility of
of Ps$_2$ formation via the Langmuir-Hinshelwood reaction
mechanism for physisorbed atoms.\cite{morisset04}

It is well established that the interaction
potential between a physisorbed atom and a
solid crystal surface is determined
by two main contributions. At large distances the
interaction is dominated by the attractive van der Waals polarization
interaction, and close to the surface the interaction is repulsive,
due in essence to the overlap of the electron wave functions of the
two subsystems.\cite{hoinkes80} At a distance $z$ from a solid
surface (the crystal occupying the lower half-space),
a physisorbed atom is subject to a potential
\begin{equation}
v(z)=v_R(z)+v_{VW}(z).\label{vz}
\end{equation}
The van der Waals interaction is written as
\begin{equation}
v_{VW}(z)=-{C\over{(z-z_{\rm vw})^3}}f(k_c(z-z_{\rm vw})),
\end{equation} \label{vdw}
with the coefficient $C$ given
by\cite{zaremba76,dzyaloshinskii61}
\begin{equation}
C={\hbar\over{4\pi}}\int_0^\infty d\xi\;\alpha(i\xi)
\left({\epsilon(i\xi)-1}\over{\epsilon(i\xi)+1}\right),
\end{equation}
where $\alpha$ is the polarizability of
the atom and $\epsilon$ is the bulk dielectric function of the
solid. The function $f$
describes the fact that the van der Waals interaction
saturates as $z$ draws closer to the reference
plane.\cite{pathak88}
The vanishing of the response at short wavelengths introduces
the cut-off wave vector $k_{\rm c}$,\cite{comment1}
for which we take the
inverse of the Ps hard-core radius, $r_{\rm c}=1.9$
$a_0$.\cite{oda01} We discuss $z_{\rm vw}$,
the so-called reference plane position, further on,
when addressing the repulsive interaction.

To obtain the dielectric function we require first to
calculate the electronic structure of the crystal. We
considered $\alpha$-quartz
(i.e., SiO$_2$ with trigonal structure, space group P3$_2$21), which
is the most common and the one used in Ref.~\onlinecite{sferlazzo85},
and have employed the experimental
structural parameters of Ref.~\onlinecite{lepage76}.
We used the highly precise all-electron
full-potential linearized augmented
plane wave (FLAPW) implementation of density-functional
theory.\cite{wimmer81} 
The band contribution to the
imaginary part of the dielectric function is calculated according
to\cite{asahi99}
\begin{eqnarray}
\varepsilon_2(\omega)={{8\pi^2e^2}\over{\Omega}}
\lim_{{\bf q}\to 0}\sum_{c,v}
\sum_{\bf k}{1\over{q^2}}
|\langle{\bf k}+{\bf q},c|e^{i{\bf q}\cdot{\bf r}}|{\bf k},v\rangle|^2
\nonumber \\
\times
\delta(\epsilon_{{\bf k}+{\bf q},c}
-\epsilon_{{\bf k},v}-\hbar\omega).
\end{eqnarray}
Here $\Omega$ is the crystal volume, and $c$ and $v$ denote the
conduction and valence bands, respectively. It is 
important to note that we use the screened-exchange local
density approximation (sX-LDA) to the exchange-correlation
potential.
This approximation is known to give excellent results regarding
the band gaps and optical properties of
$sp$-semiconductors.\cite{asahi99} This is
because of its superior description of the exchange-correlation
hole--whose long range behavior is critical in semiconductors--compared
to the well-known local density
approximation (LDA).\cite{engel97} Indeed,
the sX-LDA band gap in SiO$_2$, $E_{\rm g}=8.8$ eV, agrees very well
with the measured value of 8.9 eV,\cite{distefano71} while the LDA band
gap is only of 6.1 eV.

\begin{figure}
\includegraphics[width=0.9\hsize]{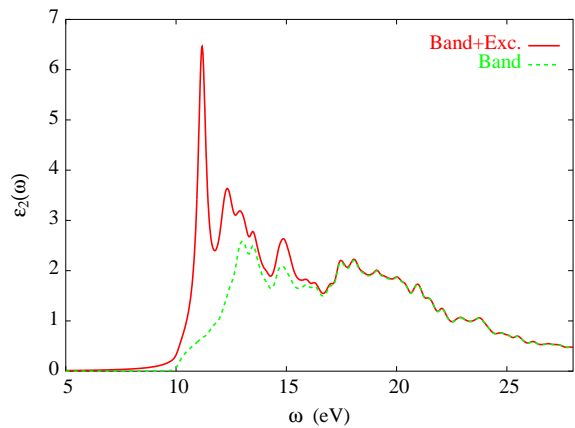}
\caption{\label{fig1} Imaginary part of the dielectric function of
$\alpha$-quartz for a polarization vector perpendicular to the
$c$-axis. The continuous line shows the total dielectric function,
arising from interband and exciton excitations. To illustrate the
excitonic effects, the dotted line shows the interband contribution
only.}
\end{figure}

Now, a recent {\it ab initio} study of excitonic effects in
$\alpha$-quartz has shown that these are crucial to understand
the structure of its optical spectrum,\cite{chang00} most
notably the strong peaks observed
near the absorption edge.\cite{phillip85}
We parametrize these contributions with a Hopfield term of the
form
$\sum_j\beta_j/(\omega_j^2-\omega^2-i\gamma\omega)$. In
Fig.~\ref{fig1} we show both the band contribution and the total
$\varepsilon_2(\omega)$ for light polarized perpendicular to the $c$-axis
of the crystal ($\alpha$-quartz is optically uniaxial). Our result
compares well with the
aforementioned works.\cite{comment2} The real part of
the dielectric function, obtained by the Kramers-Kronig
relation, yields a calculated macroscopic dielectric
constant $\varepsilon_\infty=2.38$, which is in fact the
experimental value.\cite{phillip85} The dielectric function
for imaginary argument is obtained by analytic
continuation.\cite{dzyaloshinskii61}
Our result for the van der Waals
coefficient is finally $C=13.87$ eV $a_0^3$.

\begin{figure}
\includegraphics[width=0.9\hsize]{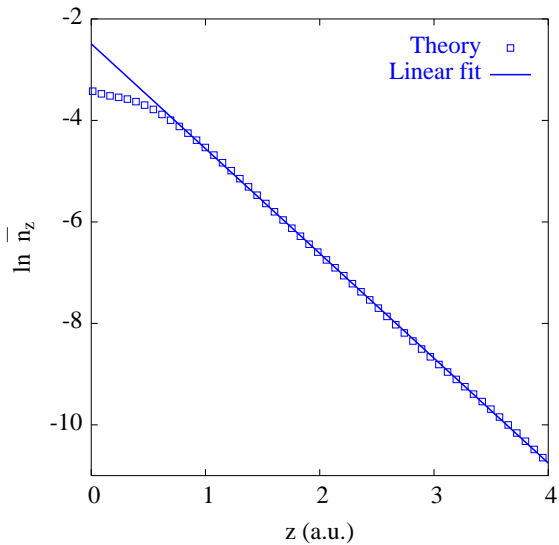}
\caption{\label{fig2} Behaviour of the electron density at the
surface of a $\alpha$-quartz slab exposing a (011) surface,
illustrated by a logarithmic plot of the
electron density averaged over the section of the unit cell,
as a function
of the coordinate $z$ perpendicular to the slab.}
\end{figure}

The polarizability of Ps is readily obtained by
(reduced) mass-rescaling that
of hydrogen. For the latter, we take the parametrization
of Dalgarno and Victor\cite{dalgarno66}
$\alpha(\omega)=\sum_n f_n/(\omega_n^2-\omega^2),$
with the oscillator strengths and frequencies they give.

The repulsive part of the interaction is essentially
proportional to the valence electron density profile near
the surface of the crystal.\cite{esbjerg80} 
We write analytically
the repulsive potential as\cite{ford76}
\begin{equation}
v_R(z)=V_0e^{-(z-z_0)/l},
\end{equation}
where $V_0$ is the intensity of the Ps work function, $l$
is the electron density decay length, and $z_0$ is the so-called
background edge position. For $V_0$ we take the experimental value,
which is
1 eV.\cite{nagashima98} To determine the two other quantities
we calculate first the
electronic density at the surface of the crystal.
For this, we considered a single
SiO$_2$ slab in which the
crystal exposes the (011) face (the so-called AT-cut
crystal) to the Ps atom, for direct comparison with
experiment.\cite{sferlazzo85}
The slab thickness was $\sim24$ $a_0$, and the unit cell contained
6 formula units. Here, two technical points are worth mentioning.
To eliminate spurious states and to
avoid any possible dipole field effects on the top surface due to
the finite thickness of the slab,
the dangling bonds at the bottom surface of the slab were
passivated with hydrogen at optimized positions. Further, the atoms
near a surface generally reconstruct, changing their relative
positions
with respect to the infinite crystal, and possibly
affecting the electron
density profile. Thus, the atom positions of the upper half of the
slab were also relaxed and optimized.

For given $z$,
we define $\bar n_z$ as the average of the valence electron
density over a section of the unit cell parallel to the $xOy$ plane. In
Fig.~\ref{fig2} we show a plot of $\ln\bar n_z$, together with a
linear fit. The origin of coordinates corresponds
to the position of the uppermost layer of atoms (oxygen).
The behaviour
becomes clearly linear at a certain point from the origin,
defining both the electron density decay length, $l$, and the background
edge position, $z_0$.\cite{comment3}
The slope of the linear fit yields $l^{-1}=2.06$ a.u., and we find
$z_0=0.95$ a.u.
In our case, furthermore, $z_0$
also indicates the value of the reference plane
position, $z_{\rm vw}$, for the van der Waals term. Indeed, in their
work
Zaremba and Kohn\cite{zaremba76} found that
in the case of insulators $z_{\rm vw}$ is located to a good
approximation at the background edge.

\begin{figure}
\includegraphics[width=0.9\hsize]{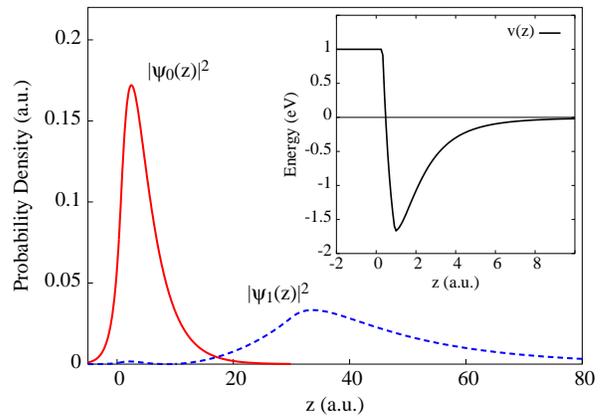}
\caption{\label{fig3} Plot of the Ps ground state probability density
(solid curve) and of its sole excited state (dotted curve). The inset
shows the interaction potential. The potential saturates once it
reaches the Ps work function value (1 eV).}
\end{figure}

Note that the total potential $v(z)$ saturates when reaching the
work function value as $z$ draws closer to the crystal. Thus, from
this point on, we take a constant potential.
Once the interaction potential is determined, we need to solve
the Schr\"odinger equation
$[-(\hbar^2/2m)\nabla^2+v(z)]\psi(z)=E\psi(z)$ and see if it admits any
bound state solution. We solve this equation semi-analytically,
exploiting an analytic expansion for $z\to+\infty$ of the form
$\exp(-|E|z)\sum_{n\geq 0} a_nz^{-n},\,\,a_n=-[Ca_{n-2}+n(n-1)a_{n-1}]/
2|E|n$. We find two bound states:
a ground state $\psi_0$ with
energy $E_0=-0.14$ eV and a single excited state $\psi_1$ with
energy $E_1=-5$ meV. In
Fig.~\ref{fig3}, we show a plot of the probability density for both
states,
as well as of the potential $v(z)$ in the inset.
Thus, the $E_0$ value we find is
remarkably close to the $-0.15$ eV estimate by
Sferlazzo and co-workers.\cite{sferlazzo85,comment4}

Given that there is no previous theoretical study along the above lines
of the problem we are considering, and that there is little
experimental information to date, it is useful to compare our
results with some known facts regarding physisorption. First, we
point out that there is an empirical,
universal relationship between the van der Waals constant $C$, the
macroscopic dielectric constant of the crystal, and the static atomic
polarizability, put forward by Hoinkes,\cite{hoinkes80} which is
expected to hold within a spread of roughly 30\%. The relation is
$C=K\alpha(\varepsilon-1)/(\varepsilon+1)$, where $K=1.41$ eV. With
$\alpha=4.5$ a.u. (static Ps polarizability) and
$\varepsilon=2.38$ one finds
$C=20.7$ eV $a_0^3$, so that our value is 33\% below. It is
notable that the Hoinkes law still applies relatively well to such a
light atom as Ps. Second, it is of interest to compare the energies we
find with those of other systems. If we take H, which is the lightest
normal atom, we find that its binding energy on graphite,
for example, is
32 meV (for other systems it is even lower).\cite{hoinkes80}
Thus, at first sight the ground
state energy we find for Ps appears to be rather large. One can
understand this, however, noting that the well depth in our case is
$\sim1.7$ eV, while in the case of H/graphite it is 43 meV.
The fact that our value of $E_0$ is such a small fraction of the well
depth is of course due to the lightness of Ps. Also, the potential
well in our case is so deep because of the strong polarizability of
Ps (8 times that of H) and its relatively small work function.
Thus, the values we find, although unusual, are overall well understood in
the framework of physisorption. They appear to be
robust also in the following sense. For comparison,
we have considered the case
of the so-called Z-cut crystal, i.e., one exposing the (001) surface to
the Ps atom.
The results are very close to those of the AT-cut crystal, with a less
than 1\% difference in the binding energies.
Furthermore, the change in energies
arising from surface reconstruction is of the same order of
magnitude.\cite{comment5}

We now briefly discuss the possibility of Ps$_2$ formation, for which
we invoke the Langmuir-Hinshelwood
mechanism.\cite{morisset04} Consider two Ps atoms initially
physisorbed on the surface of a crystal. Since the Ps atoms
attract each other with an effective Lennard-Jones potential,\cite{oda01}
they can eventually collide, recombine, and desorb. First, the fact
that the atoms are trapped at the surface increases the chance that
they do not simply rebound,\cite{morisset04} and second, the
Ps$_2$ binding energy is $E_{\rm B}=0.44$ eV. Thus, upon recombination
the Ps$_2$ molecule releases internal energy, which is transfered
to its $z$ degree of freedom, and causes the molecule to be desorbed
with an energy of $\sim0.16$ eV (the initial total energy of the two
trapped Ps atoms is 0.28 eV).

In summary, we used a first principles approach to calculate the key
parameters describing the
Ps/$\alpha$-quartz interaction potential at physisorption distances
and showed that it indeed has a bound ground state, with an energy very
close to that estimated in experiment. This result, furthermore,
is consistent with the possibility of Ps$_2$ formation at the
quartz surface via the Langmuir-Hinshelwood mechanism. We believe
our results bring strong support to the idea of Ps trapping at the
surface of certain insulators and to the fascinating prospect of
creating many-Ps systems in this way.

\begin{acknowledgments}

This work was supported by the Department of
Energy (under grant Nos. DE-FG02-88ER45372 and
DE-AC03-76SF00098/DE-FG02-07ER46352
and a computer time grant at the
National Energy Research Scientific Computing Center). We thank useful
discussions with A. P. Mills, Jr., and A. Weiss. 
B. B. benefited from the support of the Advanced Scientific Computation
Center at Northeastern University and R. S. acknowledges J.-H. Song
for useful helpful comments.

\end{acknowledgments}


\begin{thebibliography}{99}
\bibitem{comment0} We note that it is more straightforward to
discuss physisorption in the case of quartz than in Al, since in the
former case Ps is much less deformed.
\bibitem{comment2} Our spectrum presents a blue shift of
$\sim0.8$ eV compared to experiment. However, the effect of this on
the van der Waals constant is less than 0.1\%.
\bibitem{comment4}
Also of interest is that, following Ref.\onlinecite{platzman86} to  calculate the trapped Ps lifetime
change (with a calculated $\omega_p=23$ eV), we find an
increase of 1\%, i.e., close to the 2\% reported
in Ref.\onlinecite{gidley76}.
\bibitem{comment5} The main difference in this case
is in $z_0$, of the
order of 10\%, but with little effect on the binding energies.
\end{thebibliography}
\end{document}